\documentclass[10pt,twocolumn,prd,aps]{revtex4-1}
\usepackage[english]{babel}
\usepackage{amsmath}
\usepackage{amssymb}
\usepackage{graphicx}

\usepackage{hyperref}

\newcommand{\B}{\textrm{B}}
\newcommand{\vk}{\vec{k}}

\begin{document}

\title{Inflation as an amplifier: the case of Lorentz violation}

\author{Yuri Bonder}
\email{bonder@nucleares.unam.mx}
\affiliation{Instituto de Ciencias Nucleares, Universidad Nacional Aut\'onoma de M\'exico\\
Apartado Postal 70-543, Ciudad de M\'exico, 04510, M\'exico}

\author{Gabriel Le\'on}
\email{gleon@fcaglp.unlp.edu.ar}
\affiliation{Grupo de Astrof\'isica, Relatividad y Cosmolog\'ia, Facultad de Ciencias Astron\'omicas y Geof\'isicas, Universidad Nacional de La Plata\\
Paseo del Bosque S/N, 1900 La Plata, Argentina}
\affiliation{Departamento de F\'isica, Facultad de Ciencias Exactas y Naturales, Universidad de Buenos Aires, Ciudad Universitaria -
 Pab. I, 1428 Buenos Aires, Argentina.}

\begin{abstract}
Modified gravity theories are supposed to incorporate low-energy quantum-gravity effects and, at the same time, they could shed light into the dark matter and dark energy problems. Here we study a particular modification of general relativity where local Lorentz invariance is spontaneously broken and whose physical effects, despite a decade-long effort, were unknown. We show that, during inflation, this modification produces anisotropies that would generate measurable effects on the Cosmic Microwave Background. Then, by using empirical constraints on the B-mode polarization spectrum, we can estimate that the `coefficient' components absolute value have to be smaller than $10^{-43}$. This is a remarkably strong limit, in fact, it is 29 orders of magnitude better than the best constraints on similar coefficients. Thus, we propose that inflation could stringently test other modified gravity theories.
\end{abstract}

\maketitle

The quest for a gravity theory that is compatible with quantum mechanics and that, at the same time, can explain the nature of dark matter and dark energy, has lead to consider modified theories of gravity \cite{ModGrav}. These modifications come in very different forms; the common feature is that they are supposed to provide a better description of Nature and to produce small effects in regimes where the current theories have been tested. In particular, these modified theories should describe the Universe evolution from the onset of inflation since this epoch is usually assumed to be correctly described by general relativity.

Amongst the majority of cosmologists, inflation, an early era in which the Universe underwent an accelerated expansion, is held as an essential part of the standard $\Lambda$CDM cosmological model. Historically, it was conceived to solve the flatness and horizon problems of the standard Big Bang model. However, its current success is based on the power to explain the primordial inhomogeneities generation that represent the seeds of cosmic structure \cite{mukhanov1,mukhanov2,starobinsky,guth,hawking}. Furthermore, the latest Planck satellite data release indicates that inflation correctly characterizes the early Universe \cite{planck2015,planck2015likelihoods,planck2015inflation}. In particular, this data suggests that the primordial perturbations spectrum is essentially scale invariant, favoring the simplest inflationary models \cite{planck2015inflation,Martin2013}. 

In this work, we study, during the inflationary regime, a modified gravity theory that violates local Lorentz invariance. Recall that local Lorentz invariance is one of the basic tenets of general relativity and it states that there are no preferred spacetime directions. Moreover, our main motivation for considering Lorentz violation relies on studies, within prominent quantum gravity candidates, that argue that Lorentz violation may occur at the quantum gravity regime (see, e.g., Refs.~\onlinecite{KosteleckySamuel,GambiniPullin}).

A systematic program to look for Lorentz violation revolves around the general parametrization known as the Standard Model Extension (SME) \cite{SME1,SME2,Kostelecky2004}. Remarkably, this program has led to significant bounds on many parameters \cite{DataTables}. The SME is constructed in the effective field theory framework, and thus, it includes all Lorentz violating extensions to conventional physics. In particular, the SME contains a gravitational sector whose dominant correction is described by the action term \cite{Kostelecky2004}
\begin{equation}\label{mgSME action}
 S_{\rm LV} = \int d^4 x\sqrt{-g}\left(-u R + s_{ab}{R^{(T)}}^{ab} + t_{abcd}W^{abcd}\right),
\end{equation}
where $u$, $s_{ab}$, and $t_{abcd}$ are the corresponding `coefficients' that parametrize the deviation from conventional physics, and $R$, ${R^{(T)}}^{ab}$, and $W^{abcd}$ are, respectively, the curvature scalar, the traceless Ricci tensor, and the Weyl tensor. Also, $g$ is the determinant of the components of the metric $g_{ab}$. Note that we follow the notation and conventions of Ref.~\onlinecite{Wald} and we work in natural units. The coefficient $t_{abcd}$ has the index symmetries of the Weyl tensor, thus, it is completely traceless and, in four spacetime dimensions, it has $10$ independent components. 

Remarkably, before this work, the physical effects of $t_{abcd}$ were unknown \cite{BaileyKostelecky2006,tpuzz2,AltschulBaileyKostelecky2010,tpuzz3,tpuzz4,tpuzz5,Bonder2015} as all terms containing $t_{abcd}$ cancel out when the phenomenological approximations are applied; this is known as the $t$ puzzle \cite{BaileyKosteleckyXu2015}. One of the approximations that has been repeatedly used when looking for the effects of $t_{abcd}$ is to incorporate gravity as perturbations on top of a flat spacetime. In contrast, here we work in the cosmological context (cf. Ref. \onlinecite{Lambiase}) and we show that, in this setting, $t_{abcd}$ produces physical effects. In addition, we find that inflation magnifies the effects of such a term allowing us to set remarkably strong bounds on it. We believe that this result suggests that inflation can magnify the effects of other modified theories of gravity, and thus, it could be used to stringently test such theories.

It turns out that, when spacetime is dynamical, severe restrictions on the coefficients arise from the Bianchi identity \cite{Kostelecky2004}. Therefore, it is customary to assume that any Lorentz violation arises spontaneously. Now, in previous attempts to study the effects of $t_{abcd}$, its action terms were not explicitly chosen and, instead, consistency conditions fix its form perturbatively \cite{BaileyKostelecky2006}. Here, we specify such an action to be
\begin{equation}
 S_t =\int d^4x \sqrt{-g}\left(\frac{1}{2}\nabla_a t_{bcde}\nabla^a t^{bcde} -V_t\right),
 \end{equation}
with $V_t= (\kappa/2)(t^2-b^2)^2$. In addition, $t^2=t_{abcd}t^{abcd}$ and the free nonnegative parameters of the model are $\kappa $ and $b$. The former action has a conventional kinetic energy term and a Mexican hat potential for $t^2$ that produces the spontaneous Lorentz violation. Also, $\kappa$ is assumed to be large to dominate over the kinetic energy term. Furthermore, we set $u=s_{ab}=0$ because these coefficients can be moved to the matter SME sectors by metric redefinitions \cite{Bonder2015}. Thus, the total action of the model is
\begin{equation}
 S=S_{\rm EH}+S_{\rm LV}+ 16 \pi(S_{\rm M}+S_t),
 \end{equation}
where $S_{\rm EH}$ is the Einstein-Hilbert action and $S_{\rm M}$ is the matter fields action. The equation of motion associated with the metric and $t_{abcd}$ variations are, respectively,
\begin{eqnarray}
G_{ab} &=& \frac{1}{2}g_{ab} t_{cdef}W^{cdef}-6 t_{cde(a} {W_{b)}}^{cde} + 3 t_{cabd}R^{cd}\nonumber \\
&&+2\nabla^c \nabla^d t_{c(ab)d}+ 8\pi T^{({\rm M})}_{ab}+ 8\pi T^{(t)}_{ab},\\
W_{abcd}&=&16\pi \left[g^{ef}\nabla_e \nabla_f t_{abcd}+2\kappa \left(t^2-b^2\right)t_{abcd}\right],\label{t eom}
\end{eqnarray}
where the energy-momentum tensors of the matter and $t_{abcd}$, $T^{({\rm M})}_{ab}$ and $T^{(t)}_{ab}$, are defined in the standard way. In addition, there are matter fields equations of motion.

We assume that inflation can be described as a de Sitter background and that it is driven by a scalar field $\phi$ known as the inflaton. This is the only matter field we consider. We analyze two perturbations over this background: the inhomogeneous and anisotropic perturbations of standard cosmology, generated by the inhomogeneities of the inflaton, and the effects due to $t_{abcd}$. The fact that $t_{abcd}$ can be treated as a perturbation can be naively justified by the lack of empirical evidence of Lorentz violation and it becomes evident by the limits set on its values at the end of the paper. We should stress that the perturbation analysis on $t_{abcd}$, to make sense, must be considered as a perturbative expansion on $b$ since, otherwise, it is inconsistent to assume that $t_{abcd}$ lies at the bottom of the potential while it is simultaneously small. Importantly, to first order in the perturbative analysis, these two perturbations can be treated independently. Also, for simplicity, we only study the case where the effects introduced by $t_{abcd}$ are homogeneous (but anisotropic).

We use the standard notation where quantities with a bar refer to the background and the perturbations, except $t_{abcd}$, are preceded by a $\delta$. For example, $g_{ab}= \bar{g}_{ab} + \delta g_{ab}$, where the background metric, in conformal time $\eta$, has components $\bar{g}_{\mu\nu}=a^2(\eta) \eta_{\mu\nu}$, where $\eta_{\mu\nu}$ are the components of the Minkowski metric in standard coordinates and $a(\eta)$ is the scale factor and it is given by $a(\eta) = -1/(H\eta)^{(1+\epsilon)}$. The parameter $\epsilon$ is the slow roll parameter, which, during inflation, satisfies $\epsilon \ll 1$. Note that, if $\epsilon =0$, then the background is an exact de Sitter spacetime. Here $H$ denotes the Hubble factor that is essentially constant and it relates to the inflaton potential, $V$, through $H^2 = (8 \pi/3) V$. The conformal time $\eta$ is strictly negative and it runs from a largely negative quantity to zero; the value $\eta=0$ does not correspond to the inflationary period but it belongs to the radiation-dominated epoch.

At leading order in our perturbative scheme we get the Einstein equations for $\bar{g}_{ab}$ and, since $\bar{g}_{ab}$ is conformally flat, Eq.~\eqref{t eom} is identically satisfied. To first order in the perturbations we get
\begin{subequations}\label{first order eq}\begin{equation}
\delta G_{ab} = \bar{g}^{ce}\bar{g}^{df}\left(3t_{cabd} \bar{R}_{ef} + 2 \bar{\nabla}_e \bar{\nabla}_f t_{c(ab)d}\right)+8\pi\delta T_{ab}^{({\rm M})},
\end{equation}
\begin{equation}
\delta W_{abcd}=16 \pi\bar{g}^{ef}\bar{\nabla}_e \bar{\nabla}_f t_{abcd}.
\end{equation}
\end{subequations}
Note that, even though $\kappa$ is large, we neglect $\kappa (t^2- b^2)$ since we focus on initial conditions where this term vanishes and the energetic cost of making this term nonzero is controlled by $\kappa$.

We follow the convention where the index $0$ represents the conformal time and the Latin indices $i$, $j$, $k$, and $l$ stand for the spatial directions. We also use the standard scalar-vector-tensor decomposition of the metric perturbations \cite{mukhanov92}. In particular,
\begin{equation}
\delta g_{ij} = a^2 (\eta) (A \delta_{ij} + \partial_i \partial_j B + \partial_j C_i + \partial_i C_j + D_{ij} ),
\end{equation}
where $A$, $B$, $C_i$ and $D_{ij}$ are the corresponding scalar, vector and tensor perturbations. Note that $D_{ij}$ is traceless, symmetric and, in addition, it is gauge invariant. Furthermore, under the assumption that $t_{abcd}$ is homogeneous, only $D_{ij}$ is sensitive to the presence of $t_{abcd}$. Therefore, $D_{ij}$ has two contributions: a part caused by the inflaton, $D_{ij}^{(\phi)} (\vec x,\eta)$, and a homogeneous part $D_{ij} (\eta)$ due to $t_{abcd}$. Since there are no scalar perturbations associated to $t_{abcd}$, the inflaton equations of motion, for both the background and the scalar inhomogeneous perturbations, coincide with those of standard cosmology.

Moreover, mimicking the Weyl tensor decomposition \cite{Hall}, the ten independent components of $t_{abcd}$ can be split into two $3\times 3$ traceless and symmetric matrices, $t_{0i0j}$ and ${\epsilon_i}^{kl}t_{0jkl}$, where $\epsilon_{ikl}$ are the components of the volume $3$-form. In the limit under consideration, Eqs.~\eqref{first order eq} take the simple form
\begin{subequations}\label{eqsmot1}\begin{eqnarray}
a^5 D_{ij}''+ 2 a^4 a' D_{ij}' =-4 a t_{0i0j}''&+&16a' t_{0i0j}'\\
 &&+\left( 16a''-\frac{40 a'^2}{a}\right)t_{0i0j} ,\nonumber
 \end{eqnarray}
 \begin{equation}
 a^6 D_{ij}''=64 \pi\left[ a^2 t_{0i0j}'' -6 a a't_{0i0j}'+\left(6 a'^2 -4 a a''\right)t_{0i0j}\right],
\end{equation}
\end{subequations}
where the prime denotes the conformal-time derivative. Note that the equations decouple in the sense that, for given values of $i$ and $j$, only $D_{ij}$ and $t_{0i0j}$ depend on each other. Also, there are no conditions on $t_{0ijk}$, making our analysis insensitive those components.

Surprisingly, Eqs.~\eqref{eqsmot1} can be solved analytically by decoupling $t_{0i0j}$ from $D_{ij}$ after taking an additional $\eta$ derivative. We present its solutions for concrete initial conditions given at $\eta=\eta_0$ and corresponding to the onset of inflation. We take $t_{abcd}(\eta_0)$ such that $t^2(\eta_0)=b^2$, i.e., it lies at the bottom of the potential $V_t$. Furthermore, $t_{abcd}(\eta_0)$ selects a preferred spatial direction along the coordinate $x^1$, while maintaining isotropy in the $x^2-x^3$ plane. Observe that no generality is lost by choosing the spatial coordinates this way, however there are more general situations where no isotropic subspaces are left. Also, $t_{0i0j}'(\eta_0)= D_{ij}(\eta_0)=D_{ij}'(\eta_0)=0$. Note that these initial conditions on $D_{ij}$ are such that the Bunch-Davies vacuum, which is associated with the inhomogeneous tensor modes $D_{ij}^{(\phi)}$, is unperturbed. Then, the solutions of Eqs.~\eqref{eqsmot1} are of the form
\begin{eqnarray}
 t_{0i0j} (\eta)&=& C^{(1)}_{ij} |\eta|^{r_{-}} + C^{(2)}_{ij} |\eta|^{-1} + C^{(3)}_{ij} |\eta|^{r_{+}},\\
 D_{ij} (\eta) &=& K^{(1)}_{ij} |\eta|^{s_{-}} + K^{(2)}_{ij} |\eta|^3 + K^{(3)}_{ij} |\eta|^{s_{+}},
 \label{dij}
\end{eqnarray} 
where $C^{(n)}_{ij}$ and $K^{(m)}_{ij}$ are constants fixed by the initial conditions and $ r_{\pm} $, $s_{\pm}$ are known numerical factors, in fact $r_{+},s_{+} >0 $ and $r_{-},s_{-} < 0 $. It should be mentioned that it is possible find initial data for which $D_{ij}$ does not grow substantially; however, these solutions require to fine tune the initial data and are incompatible with the Bunch-Davies vacuum.
\begin{figure}
\begin{center}
\includegraphics{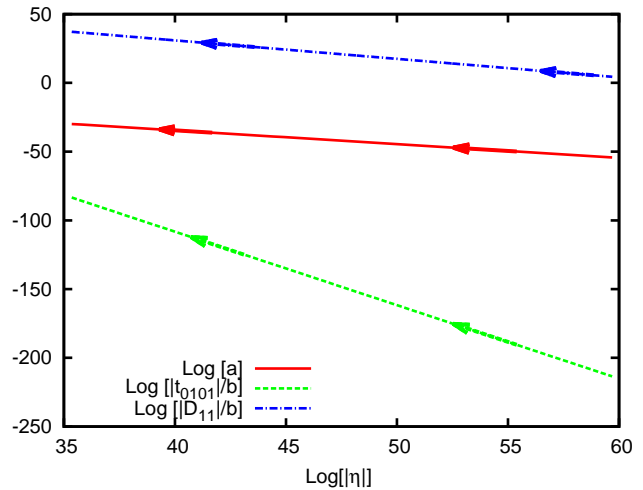}
\end{center}
\caption{Log-Log plot of the scale factor $a$ (red line), the amplitude of the components $|t_{0101}|$ (green dashed line), and $|D_{11}|$ (blue dot-dashed line), the last two quantities normalized by $b$, as functions of $|\eta|$. Inflation evolves from right to left, as indicated by the arrows.}\label{plot1}
\end{figure}

In Fig.~\ref{plot1} we plot $|t_{0i0j}|/b$, $|D_{ij}|/b$, and $a$ as functions of $\eta$ in the case where the only nonvanishing component is $i=1$, $j=1$. Observe that, due to numerical limitations, the plot begins  7 e-folds after the beginning of inflation, which in conformal time corresponds to  $\eta \sim e^{-7} \eta_0$. As it can be seen in the plot, during inflation, $|t_{0101}|$ and $|D_{11}|$ grow by many orders of magnitude. Furthermore, under the approximations we use, there are only two independent components, the other component has $i=1$ and $j=2$. However, the fields behavior in this last case is analogous to that shown in Fig.~\ref{plot1}; the difference is a $O(1)$ proportionality factor that arises when setting $t^2(\eta_0)=b^2$.

The main conclusion thus far is that $t_{abcd}$ produces physical effects that are encoded in $D_{ij}$. In what follows we describe how Cosmic Microwave Background (CMB) observations allow us to set bounds on its components. First, recall that the CMB is originated at the decoupling epoch, which occurs after the end of inflation. Thus, to estimate the effects of $t_{abcd}$ on the CMB, we should calculate the evolution of $t_{0i0j}$ and $D_{ij}$ until that epoch. However, from the end of inflation and until the time when the CMB is created, the Universe continues to expand, albeit at a much slower rate, and $|D_{ij}|$ would continue to grow. Therefore, including this stage of the Universe evolution will only make our bounds stronger and we can omit this analysis without compromising our results.

It is well known that $\epsilon$ is related with $r$, the so-called tensor-to-scalar ratio, by $r=16 \epsilon$ \cite{mukhanov92} and that the latest observational bound is $r<0.12$ (95\% C.L.) \cite{planck2015inflation}. On the other hand, when a particular value of $\epsilon$ is assumed, the inflation energy scale is automatically fixed \cite{liddle93}. Here we use $\epsilon = 10^{-3}$ to fit the experimental value of the scalar spectrum amplitude at $V/M_P^4 \epsilon \simeq 10^{-9}$, with $M_P=1/\sqrt{8\pi}$ the reduced Planck mass. In turn, $\epsilon$ sets the characteristic inflation energy scale at $V^{1/4} \simeq 10^{-3}\ M_P$ and the Hubble factor turns to be, approximately, $10^{-6}\ M_P$. This implies that $r\simeq 10^{-2}$. We also consider conventional numerical values for the beginning and end of inflation, namely, $|\eta_0| \simeq 10^6\ {\rm Mpc} = 10^{62}\ M_P^{-1}$ and assume that the inflationary regime ends after 65 $e$-folds, namely, at the conformal time $\eta_f = e^{-65} \eta_0$. All those assumptions imply that, at the end of inflation, $ |D_{11}| \simeq 10^{37}b$ and $|t_{0101}|\simeq 10^{-81}b$.

On the other hand, the primary contribution to the B-mode CMB polarization, at large angular scales, occurs due to primordial tensor perturbations, i.e., by gravitational waves produced during inflation \cite{seljak1996}. A detailed analysis of the primordial tensor perturbations modifications due to $t_{abcd}$ is presented in the Appendix \ref{Appendix}. The upshot of this Appendix is that the contribution of $t_{abcd}$ on the B-mode angular power spectrum is a constant extra term that goes like $10^{74}b^2$. Now, the B-mode polarization has not been detected \cite{planckdust}, and the only observational limit on its amplitude comes from the bound on $r$ \cite{planck2015inflation,PlanckBicep15}, which also sets the amplitude of the tensor power spectrum of the inflaton. Thus, to prevent the $t_{abcd}$ effects from dominating over the effects of the well-established inflaton quantum theory, we need to set $b<10^{-43}$. Furthermore, if the B-mode CMB polarization is ever detected, the value of the first term in the right-hand-side of Eq.~\eqref{bbound} will be fixed and we will be able to put a more precise bound on $b$.

Note that this bound on $b$ can be considered as a limit on the initial values of the coefficient components, namely, $|t_{0i0j}(\eta_0)|<10^{-43}$. To appreciate the power of inflation to test this modified theory one should compare our bounds with the best available limits on the other SME coefficients in the gravitational sector. It turns out that, in this sector, some $s_{ab}$ components, in a well defined frame centered at the Sun, need to be smaller than $10^{-14}$ to avoid producing gravitational Cherenkov radiation to a point where it would have been observed with cosmic rays detectors \cite{KosteleckyTasson2015}. Observe that the bounds obtained here are $29$ orders of magnitude more stringent! Other competitive bounds on $s_{ab}$ are placed with atomic gravimetry, Lunar Laser Ranging, Gravity Probe B, and binary pulsars observations (for a review see Ref.~\onlinecite{reviewbounds}).

To conclude, we want to stress that before this work, and for more than a decade, it was unknown whether $t_{abcd}$ produced physical effects. Here we show that it is actually physical and, by studying the effects of this coefficient in the inflationary regime, we are able to set remarkably strong bounds on its components. Perhaps, through similar analyses, other modified theories of gravity can find that unconventional physics effects get amplified, converting this type of studies into a benchmark test for such modified theories.

\begin{acknowledgments}
We acknowledge useful discussions with Quentin Bailey, Pedro Ca\~nate, Daniel Sudarsky, and Alan Kosteleck\'y, and financial support from UNAM-DGAPA-PAPIIT Grant No.~IA101116 (YB), Red FAE CONACyT (YB), and CONICET Argentina (GL). 
\end{acknowledgments}

\appendix

\section{Calculation of the $t_{abcd}$ effects on the CMB}\label{Appendix}

To relate the amplitude $|D_{ij}|$ due to $t_{abcd}$ with the observational data, we note that the complete tensor perturbation is composed by 
\begin{equation}
  D_{ij} (\vec x,\eta) = D_{ij}^{(\phi)} (\vec x,\eta) + D_{ij} (\eta).
\end{equation} 
Therefore, we can expand the complete tensor perturbation in Fourier modes, bearing in mind that the zero mode $D_{ij} (0,\eta)$ is caused by $t_{abcd}$, i.e., $D_{ij} (0,\eta)$ is the solution in Eq. \eqref{dij}, and therefore
\begin{equation}
  D_{ij} (\vec x, \eta) = \frac{1}{L^3} \sum_{\vec k}  e^{i \vec k \cdot \vec x} D_{ij} (\vec k , \eta).
\end{equation} 
Observe that we use a Fourier series instead of a Fourier integral since it is more convenient for singling out the zero mode. Formally, this corresponds to considering a cubic region of the Universe, with comoving volume $L^3$, and assuming periodic boundary conditions. At the end of the calculation we take the continuum limit $L \to \infty$. The full tensor perturbation is then 
\begin{equation}
  D_{ij}  (\vec x,\eta) = \frac{D_{ij} (0,\eta)}{L^3} +  \frac{1}{L^3} \sum_{\vec k \neq 0}  e^{i \vec k \cdot \vec x} D_{ij}^{(\phi)} (\vec k , \eta).
\end{equation} 
Furthermore, the Fourier mode $D_{ij}^{(\phi)} (\vec k,\eta)$ can be expressed as
\begin{equation}
  D_{ij}^{(\phi)} (\vec k,\eta) = \sum_{\lambda = \pm 2} \mathcal{E}_{ij} (\hat k ,\lambda) D^{(\phi)}(\vec k,\lambda,\eta),
\end{equation} 
where $\mathcal{E}_{ij}$ represents a time independent polarization tensor, $\lambda$ is the helicity, and $D^{(\phi)}(\vec k,\lambda,\eta)$ are scalar functions associated with the amplitude of the tensor power spectrum corresponding to the inflaton. From now on, we neglect the helicity since it only contributes by a factor of $2$ to the tensor power spectrum of the inflaton.  

Now, the B-mode polarization can be decomposed in spherical harmonics \cite{zaldarriaga1996,seljak1996}; the expansion coefficients $a_{lm}^\B$ are given by
\begin{equation}\label{almB}
  a_{lm}^\B = \frac{4\pi (-i)^l}{L^3} \sum_{\vec k} \Delta_{\B,l} (k) D(\vec k, \eta_f) Y_{lm} (\hat k)^*.
\end{equation} 
The latter expression includes the zero mode $D(0, \eta_f)$, which we specify later. The transfer functions $\Delta_{\B,l} (k)$ encode all the physics from the end of the inflationary era to the time of decoupling. Given that we are neglecting the effects of post-inflationary physics, and we are only interested in the amplitude rather than the shape of the spectrum, we take $\Delta_{\textrm{B}, l} (k) \simeq j_l (k R_D)$, with $j_l$ the spherical Bessel functions and $R_D$ the comoving radius of the last scattering surface. Note that, for the zero mode, $j_l (0)$ vanishes for all $l$ except for $l=0$, where $j_0(0) = 1$, and $Y_{00}$ is real and independent of $\hat k$.

Moreover, the B-mode polarization data are presented in terms of the B-mode angular power spectrum defined as
\begin{equation}\label{defClBB}
  C_{l}^{\B \B} = \frac{1}{2l+1} \sum_m \langle a_{lm}^\B a_{lm}^{\B *} \rangle,
\end{equation} 
where $\langle \cdot \rangle$ denotes ensemble average. From Eq. \eqref{almB}, we have
\begin{eqnarray}\label{almB2}
  \langle a_{lm}^\B a_{lm}^{\B *} \rangle &=&  \frac{16 \pi^2}{L^6} \sum_{\vec k,\vec k'}   \Delta_{\B,l} (k) \Delta_{\B,l} (k')  \nonumber \\
  &\times& \langle  D(\vec k, \eta_f) D(\vec k', \eta_f)^* \rangle Y_{lm} (\hat k)^* Y_{lm} (\hat k').
\end{eqnarray} 
The nature of the zero mode $D(0,\eta_f)$ and the rest of the modes $D^{(\phi)} (\vec k,\eta_f)$ is different; the former is a classical scalar field associated with the SME while the latter is a classical stochastic field coming from the quantum inflaton fluctuations. Henceforth, the ensemble average $ \langle  D(\vec k, \eta_f) D(\vec k', \eta_f)^* \rangle $ can be decomposed in $4$ terms: $ \langle  D  (0, \eta_f) D (0, \eta_f)^* \rangle $,    $ \langle  D^{(\phi)}  (\vec k, \eta_f) D^{(\phi)} (\vec k', \eta_f)^* \rangle $,  $  D (0, \eta_f) \langle  D^{(\phi)} (\vec k', \eta_f)^* \rangle $, and  $ \langle  D^{(\phi)}  (\vec k, \eta_f) \rangle  D(0, \eta_f)^* $. The last two terms vanish since  $ \langle  D^{(\phi)}  (\vec k, \eta_f) \rangle = 0$, while the first term is simply $|D (0,\eta_f)|^2$. The remaining term is associated with the dimensionless tensor power spectrum of the tensor metric perturbations corresponding to the inflaton, namely 
\begin{equation}\label{PT}
  \langle  D^{(\phi)}  (\vec k, \eta_f) D^{(\phi)} (\vec k', \eta_f)^* \rangle = \frac{2\pi^2}{k^3} P_T (k,\eta_f) L^3 \delta_{\vec k,\vec k'}.
\end{equation} 

In the continuum limit
\begin{equation}\label{ampzeromode}
 \frac{|D (0,\eta_f)|^2}{2\pi^2 L^3}  \to |D_{ij}(\eta_f) D^{ij} (\eta_f)| 
\end{equation} 
with $D_{ij}$ the solution given in Eq. \eqref{dij}. Hence, the expression for the ensemble average is
\begin{equation}\label{avg}
  \langle  D(\vec k, \eta_f) D(\vec k', \eta_f)^* \rangle = |D(k,\eta_f)|^2 \frac{L^3}{(2 \pi)^3} \delta_{\vec k , \vec k'}.
\end{equation}
If $k=0$, then $|D(k,\eta_f)|^2 $ is given by Eq. \eqref{ampzeromode}; if $k\neq0$, then $|D(k,\eta_f)|^2 = ({2\pi^2}/{k^3}) P_T (k,\eta_f) $. Inserting the expression for the ensemble average, Eq. \eqref{avg}, into Eq. \eqref{almB2}, and summing over $\vec k'$ yields
\begin{equation}
  \langle a_{lm}^\B a_{lm}^{\B *} \rangle =  \frac{2}{\pi L^3} \sum_{\vk} j_l(kR_D)^2 |Y_{lm} (\hat k)|^2 |D(k,\eta_f)|^2.
\end{equation} 
Consequently, using the definition \eqref{defClBB}, the B-mode angular spectrum is given by
\begin{equation}
  C_l^{\B \B} =  \frac{1}{2 \pi^2 L^3} \sum_{\vk} j_l(kR_D)^2  |D(k,\eta_f)|^2,
\end{equation} 
where we used the identity $\sum_m |Y_{lm} (\hat k)|^2 = (2l+1)/4\pi$. Moreover, separating the zero mode from the rest of the modes results in
\begin{equation}
  C_l^{\B \B} =  \frac{|D (0,\eta_f)|^2}{2\pi^2 L^3} + \frac{1}{L^3} \sum_{\vk \neq 0} \frac{j_l(kR_D)^2}{k^3} P_T (k,\eta_f).
\end{equation}  
Taking now the continuum limit $L \to \infty$, the latter expression becomes 
\begin{equation}
  C_l^{\B \B} = |D_{ij} (\eta_f)  D^{ij} (\eta_f)|  + 4\pi  \int_{0^+}^\infty dk \:  \frac{j_l(kR_D)^2}{k} P_T(k,\eta_f).
\end{equation} 
Finally, using the known expression for the (dimensionless) tensor power spectrum $P_T (k,\eta_f) \simeq H^2/M_P^2$ and our result $|D_{ij} (\eta_f)  D^{ij} (\eta_f)| \simeq 10^{74} b^2$ the B-mode angular spectrum is
\begin{equation}
  C_l^{\B \B} \simeq  10^{74} b^2 + \frac{H^2}{M_P^2} \frac{1}{l(l+1)},
\end{equation} 
with $l = 1, 2, \ldots$. In addition, the numerical values that lead to $|D_{ij} (\eta_f)  D^{ij} (\eta_f)|  \simeq 10^{74} b^2$ also imply that ${H^2}/{M_P^2} \simeq 10^{-12}$. Also, the effect of the primordial tensor perturbations on the tensor power spectrum is dominant at large angular scales, i.e., at the lowest multipoles $l \lesssim 10$. Therefore, an estimated value for the B-mode angular spectrum is
%%%%%%%%%%%hasta aqui
% Specifically, the data associated to the B-mode polarization is characterized by the BB angular power spectrum
% \begin{equation}
%  C_l^{\textrm{BB}} = (4\pi)^2 \int dk k^2 P_T (k) \Delta^2_{\textrm{B} l} (k),
% \end{equation} 
% where $k$ is the magnitude of the Fourier vector associated with the spatial coordinates, $P_T(k)$ is the primordial tensor power spectrum, and $\Delta^2_{\textrm{B} l} (k)$ are the B-mode transfer functions which encode post-inflationary physics. Since we are only interested in the angular spectrum amplitude, we can take $\Delta^2_{\textrm{B} l} (k) \simeq j_l^2(k R_D)$, with $j_l$ the spherical Bessel functions and $R_D$ the comoving radius of the last scattering surface. Additionally, the standard prediction for the tensor power spectrum is $P_T(k) \sim H^2/M_P^2 k^3$. Thus, in the conventional case, the estimated amplitude of the angular spectrum satisfies $ l(l+1)C_l^{\textrm{BB}} \simeq H^2/M_P^2 \simeq 10^{-12}$. Here, after considering the homogeneous tensor perturbation $D_{ij}$, it becomes
%%%%%%%%%%%%%%%%%%%%%%%%%%%%%%%%%%%%%%%%%%%%%%%%%%%%%%%%%%%%%%%%%%%%%%%%%%%%%%%%%%%%%
\begin{equation}\label{bbound}
 l(l+1)C_l^{\textrm{BB}} \simeq 10^{-12} + 10^{74} b^2.\end{equation}
 This is the expression we use to set limits on $b$.
 
\bibliographystyle{h-physrev}

\bibliography{SMEcosm} 

\end{document}